\documentclass[10pt,letterpaper]{article}
\usepackage{opex3}
\usepackage{graphicx}
\usepackage{mathptmx}
\usepackage{bm}
\usepackage{amssymb}
\usepackage{pifont}
\usepackage{cite}
\begin{document}

\title{Nonlinearity in Single Photon Detection: Modeling and Quantum Tomography}

\author{Mohsen K. Akhlaghi,$^{1,*}$ A. Hamed Majedi,$^{1}$ and Jeff S. Lundeen$^{2}$}

\address{$^{1}$Institute for Quantum Computing and ECE Department,University of Waterloo,
200 University Ave West, Waterloo, ON, Canada, N2L 3G1}
\address{$^{2}$National Research Council - Institute for National
Measurement Standards, 1200 Montreal Road, Ottawa, ON, Canada, K1A
0R6}

\email{$^{*}$mkeshava@maxwell.uwaterloo.ca}



\begin{abstract}
Single Photon Detectors are integral to quantum optics and quantum
information. Superconducting Nanowire based detectors exhibit new
levels of performance, but have no accepted quantum optical model
that is valid for multiple input photons. By performing Detector
Tomography, we improve the recently proposed model [M.K. Akhlaghi
and A.H. Majedi, IEEE Trans. Appl. Supercond. \textbf{19}, 361
(2009)] and also investigate the manner in which these detectors
respond nonlinearly to light, a valuable feature for some
applications. We develop a device independent model for Single
Photon Detectors that incorporates this nonlinearity.
\end{abstract}




\section{Introduction}

Single Photon Detectors (SPD) have been central to our exploration
of the fundamental limits of imaging \cite{Lugiato2002},
interferometry \cite{ Mitchell2004}, and communication
\cite{Barreiro2008}.  They are used in astronomy \cite{Daigle2009},
laser ranging \cite{Newsom2009}, and biological imaging
\cite{Betzig2006}. The SPD is a building block from which most more
complicated optical detectors can be built. As examples, a photon
number resolving detector can be built by multiplexing many SPDs
\cite{Achilles2003}, a phase sensitive detector by mixing the input
light with an auxiliary laser beam \cite{Puentes2009}, and an
entanglement (i.e. Bell-state) detector by coincidence counting
between SPDs behind a polarizing beamsplitter \cite{Michler1996}.
Consequently, SPDs have been ubiquitous in quantum optics and
quantum information experiments performed up to this day.

Throughout, they have been described by a simple model. SPDs are
binary detectors (`Click', or `No Click'), and thus 
any non-zero number of detected photons will result in the same
response: a Click. This behavior is contained within the standard
model for the SPD's positive-operator-valued measure (POVM):
\begin{equation}
{\pi}^{SPD}_0=\sum^{\infty}_{n=0}{{(1-P_1)}^n}\mid n \rangle \langle
n \mid, \label{eq:1}
\end{equation}
\noindent where ${\pi }^{SPD}_0$ is the No Click operator, ${P}_{1}$
is the quantum efficiency, $\mid n \rangle$ is an n-photon state,
and the Click operator is ${\pi }^{SPD}_1=1-{\pi }^{SPD}_0$. This
simple model neglects the optical or electrical nonlinearities that
likely exist, to some degree, in all SPDs. Nonlinear quantum optical
detectors can prepare novel quantum optical states
\cite{Ourjoumtsev2007}, characterize the photon statistics of input
states \cite{Achilles2003}, and measure the temporal profile of
optical wavepackets through intensity autocorrelation \cite{Xu2002}.
Advanced SPDs, with their sophisticated structures
\cite{Hadfield2009}, are good candidates to violate Eq.~(\ref{eq:1})
through non-independent photoemissions and/or nonlinear gain
mechanism.

A good example is Superconducting Nanowire SPDs (SNSPD), which,
compared to conventional SPDs, offer higher maximum count rate,
lower timing jitter, broader spectral range, and superior detection
efficiencies at telecom wavelengths \cite{Hadfield2009}. These
detectors consist of a meandering superconducting nanowire carrying
a constant bias current close to their critical current. A single
photon deposits enough heat to create a hotspot that is not
superconducting. With the help of the bias current this can initiate
a resistive bridge across the nanowire with some probability. The
resulting voltage spike signals the detection of a single photon
\cite{Goltsman2005, Semenov2001}. Unfortunately,
apart from such descriptions and calculations, there is no
associated full quantum optical model (i.e. valid for multiple
incident photons) from which we could formulate a general POVM for a
single element SNSPD.

In an attempt to quantitatively describe the different observed
experimental results within a single formalism, Akhlaghi and Majedi
\cite{Akhlaghi09} proposed a semi-empirical model (AM model). 
In this model, the absorption of two or more photons at one point in
the wire could result in cooperative resistive bridge formation.
Hence, one might expect a nonlinear response in SNSPDs. Moreover, it
might dominate the linear response at low bias currents. This is in
agreement with the early experimental results on characterization of
SNSPDs \cite{Goltsman2001}. The model has been shown to work well
with input light in a coherent state.

To further test this model we perform tomography on a SNSPD.
Detector Tomography \cite{Lundeen2009, Coldenstrodt2009} is an
agnostic procedure to determine the POVM of a detector. The detector
is treated as a Black Box in that we do not need to know its
mechanism or make ancillary assumptions about it. Just recently
demonstrated in principle, Detector Tomography has yet to be applied
to a detector without an accepted model for its POVM.

\section{Detector Tomography}

\begin{figure}[htbp]
\centering\includegraphics[width=9.0cm]{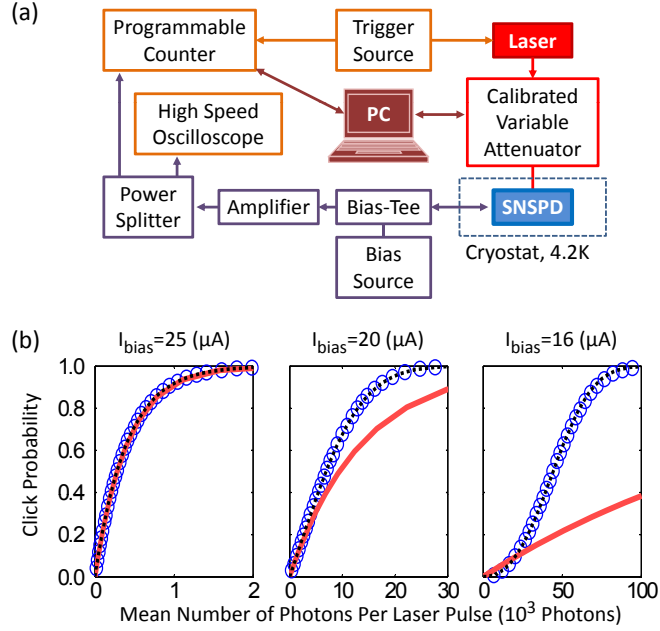}
\caption{\label{fig:1} (a) Schematics of the experimental setup. (b)
Click Probability at different bias currents. The critical current
was measured to be 26.0$\pm$0.5$\mu$A. The blue circles are measured
Click probabilities (not all points are shown). The red lines are
calculated using the linear model in Eq.~(\ref{eq:1}) by calculating
the efficiency parameter $P_1$ when the click probability is equal
to 0.1. The black dotted lines are from the nonlinear SPD Model in
Eq.~(\ref{eq:6}). $\{{P}_{n}\}$ equals $\{$7.30e-4, 2.49e-3$\}$,
$\{$9.72e-6, 7.15e-5, 8.14e-9$\}$, $\{$0, 7.33e-8, 2.87e-10,
2.81e-14$\}$ for 25, 20, and 16$\mu$A, respectively.}
\end{figure}

To perform tomography we create a coherent state with a 1310nm laser
that produces 200 ps pulses at 100 kHz. We use a programmable
attenuator to set the coherent state amplitude, $\alpha$. We
sequentially send ${R}_{T}={10}^{5}$ copies of this state into the
SNSPD system and record the number of detector Clicks, ${R}_{1}$.
This is repeated for a set of D states $\{\mid {\alpha}_{i}
\rangle\}$, increasing $\alpha$ from ${\alpha}_{0}=0$ until the
detector response is unchanging at ${\alpha}_{D}$. i.e.
$\partial{R}_{1}/\partial\alpha=0$. The estimated Click probability,
${R}_{1}/{R}_{T}$, is equal to the Q-function of the Click POVM
operator, $Q\left(\alpha\right)$. This, in itself, completely
characterizes the detector response.

Our SNSPD consists of 4 nm thick by 120 nm wide Niobium Nitride
meandering wire on sapphire. The active area is square with a fill
factor of 60\%. The SNSPD is coupled to an optical fiber and cooled
down to 4.2K by exploiting a cryogenic setup (described in
\cite{Orgiazzi2009}). Fig.~\ref{fig:1}(a) shows a schematic of the
measurement setup. The bias current is applied using a bias-Tee. The
weak response of the SNSPD after amplification is split between an
oscilloscope and a counter (see \cite{Yan2009} for more detail). We
use a 20ns counting gate, triggered by the laser, to reduce the dark
count contribution from the times between input pulses.

We perform Detector Tomography (i.e. measure Q-function) at three
bias currents, 25, 20, and 16$\mu$A. In Fig.~\ref{fig:1}(b), we plot
the measured response for each of these (blue circles). We expect a
standard linear SPD response at 25$\mu$A since this is the normal
operation mode. Using Eq.~(\ref{eq:1}), one expects a Q-function of
the form $Q(\alpha)=1-\exp \left(-P_1{|\alpha|}^2\right)$. We
estimate ${P}_{1}$ using a single data point at
$Q(\alpha)={R}_{1}/{R}_{T}=0.1$. Indeed, using this ${P}_{1}$, the
resulting predicted response (red line) agrees well with the
measured response. Repeating this analysis for 20 and 16$\mu$A, we
find that the estimated ${P}_{1}$s, and thus, quantum efficiencies
decrease as the bias current decreases. More significantly, the
disagreement between the shape of the predicted and measured Click
Probability distributions is substantial. Evidently, a SNSPD quickly
becomes nonlinear as the bias current is lowered.

A general limitation of Q-functions is that they cannot be used to
calculate a detector's response to arbitrary states of light
\cite{Leonhardt1997}. Instead, one typically uses the POVM operators
represented in the photon number basis. We find the SNSPD POVM by
fitting it to the D measured statistics ${R}_{1}/{R}_{T}$ according
to the methods given in \cite{Feito2009}. Since the detector lacks
phase sensitivity, the off-diagonal elements of the Click operator
are zero. We thus represent it as a vector ${\Pi}$ (with dimensions
${N\times1}$). We truncate $\Pi$ at a number state $N-1$ that is
sufficiently high that $\Pi(N)\approx1$.

\begin{figure}[htbp]
\centering\includegraphics[width=8.0cm]{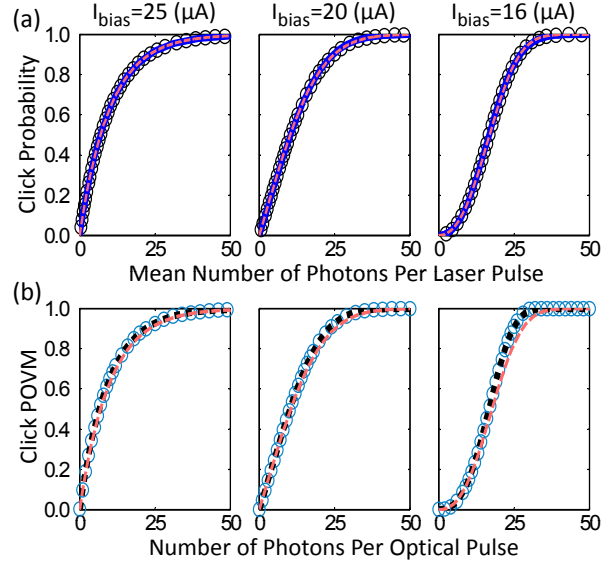}
\caption{\label{fig:2} (a) The Q-function (i.e. click probability
for coherent state inputs) of the scaled detector at different bias
currents. The raw scaled data (black circles) agrees well with the
tomographic POVM (blue line) and the AM Model (red dashed line). (b)
The corresponding Click POVM operator. The operator found from
tomography (blue circles) disagrees with that from the AM Model (red
dashed line) but agrees with that from the nonlinear SPD Model
(black dotted line). In the latter, $\{P_n\}$ equals \{7.29e-4,
9.95e-2\}, \{1.08e-5, 4.76e-2, 3.74e-3, 1.13e-4\}, \{0, 1.97e-4,
2.01e-3, 4.87e-4, 5.07e-5\} for 25, 20, and 16$\mu$A, respectively.}
\end{figure}

One difference with \cite{Feito2009} is that these particular SNSPDs
have low system quantum efficiency (0.2\% or less at 1310nm) and
thus the maximum photon number, $N-1$, required to span their
response was large. Instead of using large matrices in the fitting,
we scaled the inputs $\{\mid {\alpha}_{i} \rangle\}$ by a factor
$k\ll1$. For each bias current, $k$ is chosen so that the Click
Probability is 95\% at an average photon number $\langle n \rangle
=30$. This scaled data is shown in Fig.~\ref{fig:2}(a) (black
circles). We plot $\Pi$ determined
from it 
in Fig.~\ref{fig:2}(b) (blue circles). Using this $\Pi$, in
Fig.~\ref{fig:2}(a) (blue line) we plot the predicted detector
response to coherent input states. This fits the scaled data well,
confirming the fitting procedure.

If scaled by too large a factor the POVM will, at the very least, be
unrepresentative of the detector and, at worst, be physically
impossible. Thus we test the estimated POVM for validity by adding
$k$ loss back into the POVM (i.e.
${\Pi_{unscaled}=L^{-1}\Pi_{scaled}}$, where ${L}$ is the binomial
distribution matrix described in \cite{Feito2009}) and predicting
the detector response to the original unscaled inputs, $\{\mid
{\alpha}_{i} \rangle\}$. For all the bias currents, the difference
between the predicted click probability and the raw data (i.e.
$\left|R_{1}/R_{T}-\mathrm{Tr}{\left[\Pi_{unscaled}\left|\alpha_{i}\right\rangle
\left\langle \alpha_{i}\right|\right]}\right|$) is less than
0.15$\%$ on average and has a maximum of 1.4$\%$. This indicates
that we have accurately estimated the SNSPD POVM using the scaling
technique.

\section{Generalized Model}

The lack of ancillary assumptions and models in Detector Tomography
make it general and objective. At the same time, it provides a
surplus of information (i.e. $O\left(N\right)$ parameters) that can
be difficult to interpret. Tomography can hardly replace the natural
ease and intuition that is associated with a model. A good model
candidate is the AM model. Indeed, plotting the AM model in
Fig.~\ref{fig:2}(a) (red dashed line) shows that it can be fit to
the scaled data.

To find an analytic expression for the SNSPD POVM in the number
state basis we repeated the derivation of the AM model given in
\cite{Akhlaghi09} with one change:  We derived the detector response
to a Fock state rather than a coherent state input. The parameter
values in the model should remain the same in either case. Thus,
using the parameter values from the AM model curves in
Fig.~\ref{fig:2}(a) we plot the AM POVM in Fig.~\ref{fig:2}(b) (red
dashed line). Substantial disagreement between the tomographic POVM
and the AM POVM suggests there is an error.

Indeed, further investigation revealed that in Eqs. (4) and (5) in
\cite{Akhlaghi09} the average photon number is used in place of the
actual photon distribution. The resulting error is only revealed in
the comparison of analytical and tomographic POVMs. We fixed this
error and found that the resulting analytic POVM has a form that is
applicable to general SPD detectors. Leaving behind the SNSPD
specific derivation in \cite{Akhlaghi09}, we now introduce a broad
detector model that is able to describe nonlinear SPDs. We do this
by generalizing Eq.~(\ref{eq:1}) to multiphoton detection.

Consider a binary detector that is only sensitive to $n$ number of
photons; any less and the detector responds with No Click, any more
and it still only outputs one Click. This is the n-photon
generalization of the SPD and, hence, we call it an n-photon
detector (NPD). If $m>n$ photons impinge on the detector, there are
m choose $n$ ways for those $m$ photons to trigger the NPD.
Consequently the generalization of Eq.~(\ref{eq:1}) is:
\begin{equation} {\pi}^{NPD}_0=\sum^{\infty }_{m=0}
{{\left(1-P_n\right)}^{m \choose n}} \left|m\right\rangle
\left\langle m\right|, \label{eq:3}
\end{equation}
where ${\pi}^{NPD}_0$ is the No Click operator, ${P}_{n}$ is the
n-photon detection efficiency and ${m \choose n}$ is the binomial
coefficient ($=0$ for $n>m$, $=1$ for $n=0$). This generalization
works even for a zero photon detector. We can identify $P_0$ as what
is commonly called the `dark count probability'.

\begin{figure}[htbp]
\centering\includegraphics[width=9.0cm]{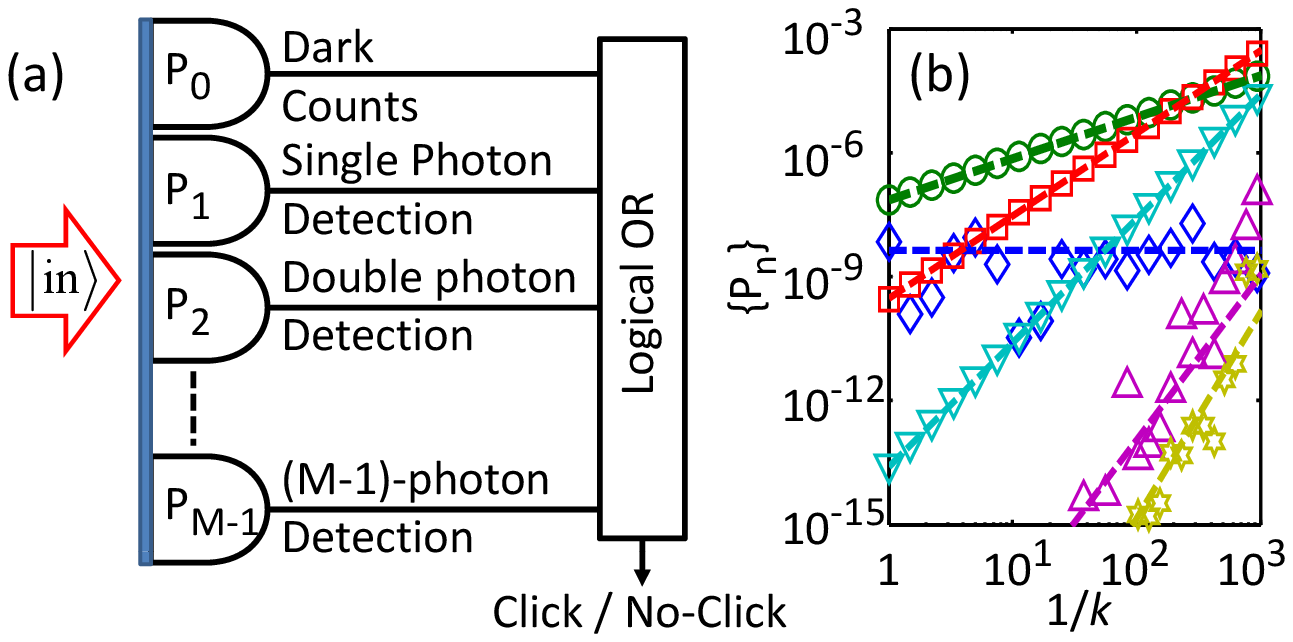}
\caption{\label{fig:3}(a) The Nonlinear Single Photon Detector
model. Each element represents an n-Photon Detector. A broadly
applicable model is created by logically ORing these elements. (b)
$\{{P}_{n}\}$ at 16$\mu$A found under different scaling factors,
$k$. $\diamondsuit$, $\bigcirc$, $\square$, $\bigtriangledown$,
$\bigtriangleup$, and \ding{73} represent ${P}_{n=0}$ to ${P}_{n=5}$
respectively. The dotted lines have slopes equal to $n$. The plot
shows a linear optical loss of $\eta$ scales $P_n\to{\eta}^nP_n$.}
\end{figure}

A nonlinear SPD can be modeled as concurrent NPDs. As shown in
Fig.~\ref{fig:3}(a), the model consists of a logical OR between M
NPDs, where M represents the maximum number of mechanisms that
should be present in order to describe the response of the detector
before it saturates at high intensities. The associated POVM
operators are:
\begin{equation}
{\pi }^{NL}_1={\rm l}-\sum^{\infty}_{m=0} {\prod^{M-1}_{n=0}{
{\left(1-P_n\right)}^{m \choose n}}}\left|m\right\rangle
\left\langle m\right|, \label{eq:5}
\end{equation}
\noindent for Click and ${\pi }^{NL}_0=1-{\pi }^{NL}_1$ for No
Click. The Click probability of the nonlinear SPD in response to a
coherent state $\mid \alpha \rangle$ is:
\begin{equation} P_{Click}{\left(  \alpha \right)}=1-
\sum^{\infty}_{m=0}{{\rm e}^{-{|\alpha |}^2}\frac{{|\alpha
|}^{2m}}{m!} \prod^{M-1}_{n=0}{\left(1-P_n\right)}^{m \choose n}}.
\label{eq:50}
\end{equation}
\noindent This can be rewritten in matrix form as:
\begin{equation}C=F{\Pi}={F}\left(E-\exp \left(GH\right)\right),
\label{eq:6}
\end{equation}
\noindent where $C$  (with dimensions ${D\times1}$), includes the D
measured statistics, ${R}_{1}/{R}_{T}$; ${\Pi}$ (${N\times1}$)
includes the diagonal elements of ${\pi}^{NL}_1$; ${F}$ contains the
D coherent state probes,
$F_{i,j}={{\left|{\alpha}_i\right|}^{2j}{\exp}
\left(-{\left|{\alpha}_i\right|}^2\right)}/j!$; ${E}$ is a matrix of
ones; ${G}$ is a matrix of binomial coefficients such that
$G_{i,j}={j-1 \choose i-1}$; and ${H}$ (${M\times1}$) is an unknown
matrix which includes the unknown set $\{P_n\}$,
$H_{i,1}=\ln(1-P_{n=i-1})$.

We estimate $H$ by solving the following constrained nonlinear
multivariable optimization problem:
\begin{equation}\min{\left\Vert\frac{C-FE+F\exp(GH)}{C}\right\Vert}_2\ ,\ \label{eq:7}
\end{equation}
\noindent subject to $H\le 0$. The second norm of a matrix is
defined as ${\left\Vert A \right\Vert}_2 = {\left( \sum_{i,j}{\left|
A_{i,j}\right|}^2 \right)}^{1/2}$. Each element of the expression is
normalized to $C$ to give equal weighting to all the points. The
constraint of the problem ensures the optimization leads to a
physical result for the set $\{P_n\}$. We note the function ${\rm
exp}(\alpha x)$ is convex on $\mathbb{R}$ for any $\alpha \in
\mathbb{R}$ \cite{Boyd2004}, which also makes Eq.~(\ref{eq:7})
convex.

We first solve Eq.~(\ref{eq:7}) for scaled input states. From the
estimated $\{P_n\}$ we only keep those elements that change the
minimum of Eq.~(\ref{eq:7}) by more than 1\% (25$\mu$A: $\{P_0,\
P_1\}$; 20$\mu$A: $\{P_0,\ \dots \ ,\ P_4\}$; 16$\mu$A: $\{P_1,\
\dots \ ,\ P_4\}$). These parameters classify the operation of the
SNSPD, from a standard SPD at 25$\mu$A to a composite of one, two,
three, and four photon detectors at 16$\mu$A.

From these $\{P_n\}$ we calculate the nonlinear SPD Click POVM
operator for the three bias currents in Fig.~\ref{fig:2}(b) (black
dotted line). They agree with the tomography POVMs to within 1\% for
most of the elements of $\Pi$. For 25$\mu$A and 20$\mu$A the maximum
difference is 3\% at ${\Pi}\left(n=1\right)$ and 6\% at
${\Pi}\left(n=3\right)$ for 16$\mu$A. This excludes the large error
at ${\Pi}\left(n=0\right)$, which we attribute to insufficient
measured statistics at extremely small mean photon numbers. The
Quantum Fidelity (see \cite{Feito2009} for a definition) of the
model and tomography operators are above 99.8\% for all three bias
currents. Thus, the model successfully gives the POVM of the SNSPDs
including their nonlinearity, but with a dramatic reduction in the
number of parameters compared to tomography.

We can attribute the SNSPD nonlinearity to the fact that at lower
bias currents 
the absorption of multiple photons in close proximity and within a
short period of time can inject enough energy to switch the wire to
normal state more efficiently than independent absorptions.

In Fig.~\ref{fig:1}(b), we plot the model's predicted response to
the coherent state inputs, with parameters from the unscaled data
(black dotted line). The difference with the measured statistics is
less than 0.34\% on average (i.e the minimum found in
Eq.~(\ref{eq:7}) divided by the number of points).

Unlike the standard SPD, where $P_1\rightarrow \eta P_1$ under a
preceding optical loss of $\eta$, there is no analytic formulae for
how $\{P_n\}$ transform under loss. By inspection of the scaled and
non-scaled model fits (see Fig.~\ref{fig:3}(b)), however, each
element of $\{P_n\}$ that is significant approximately scales as
${\eta}^n$. So removing any linear optical inefficiency from a
nonlinear photon counter makes it more nonlinear. We also note $P_0$
is not dependent on the optical input as expected.



\section{Conclusion}
The almost identical outcome of Detector Tomography and the
nonlinear SPD model confirms both of them are reliable. This is a
good example of how Detector Tomography is particularly useful for
characterizing detectors outside their normal operating regime,
where there is no model for their operation. We expect the nonlinear
SNSPD model to be useful for other nonlinear binary detectors such
as two-photon absorbing Avalanche Photodiodes and Electron
Multiplying CCDs (thresh-holded). It will also be useful for
characterizing conventional SPDs and putting limits on their
nonlinearity.

\section*{Acknowledgments}
Recently, a preprint of a related paper has appeared
\cite{Brida2011}. We acknowledge the financial support of NSERC and
BDC. We also acknowledge Jean-Luc F. Orgiazzi for development of the
cryogenic setup.

\end{document}